\documentclass{aps2010} 

\usepackage{amsmath}
\usepackage[numbers]{natbib}
\usepackage[dvips]{graphicx}

\title {Computationally efficient algorithm for fast transients detection}

\author[GS]{\textbf{\emph{Gene Soudlenkov}}}
\author[SK]{\textbf{\emph{Vyacheslav V. Kitaev}}}

\address[GS]{School of Engineering, Auckland University of Technology, Auckland, 1142, New Zealand; xpd2710@aut.ac.nz}
\address[SK]{School of Engineering, Auckland University of Technology, Auckland, 1142, New Zealand; slava.kitaev@aut.ac.nz}

\begin{document}

\maketitleblock  

\label{firstpage}

\begin{abstract}
Computationally inexpensive algorithm for detecting of dispersed transients has been developed using Cumulative Sums (CUSUM) scheme for detecting abrupt changes in statistical 
characteristics of the signal. The efficiency of the algorithm is demonstrated on pulsar PSR J0835-4510. 
\end{abstract}

\section{Introduction}
Transient radio sky or time-variable radio sources in space have been recognized 
as one of the key science drivers for the Square Kilometre Array (SKA) \citep{b26}
what will be reflected in its design and operations \citep{b11}.
    This science area is marked as “Exploration of the Unknown” what includes the likely 
    discovery of new classes of objects and phenomena. Yet the statistical properties of transient radio sky remain largely unknown, 
    what includes even the high-energy transients as they seem to be unfolding on very short time scales.

In detection approach for the arrays such as MWA, LOFAR and SKA we propose to acknowledge the need for reducing data transport and computational cost while 
searching for fast transients. Therefore, we suggest that such detection needs to be done in stages with reduced data for storing and transporting at every consequent stage. 
Firstly, the presence of the abrupt change in raw time-domain data needs to be identified. Secondly, the evidences of the extraterrestrial
nature of the detected signal need to be found. Thirdly, if such evidence is present, more detailed analysis can be applied. In this paper we present an attempt to develop a reliable and cost-effective technique for the first stage of detection using statistical signal detection approach.

\section{Detection approach}

The proposed method of handling received time domain sequences is based on online abrupt changes detection scheme. 
The algorithm operates in time domain and is required to identify the onset of the signal by checking if the threshold 
of the detection process has been reached. 

Let’s consider the signal with the following structure:
\begin{equation}\label{eq2}
\left\{\begin{array}{l} {H_{0} :X(t)=n (t)} \\ {H_{1} :X(t)=s(t)  +n (t)} \end{array}\right.
\end{equation}
Where \textit{t} is time, \textit{n} (\textit{t}) $\in \mathcal{N}(\mu_0,\sigma_0^2)$ is a Gaussian noise process with mean $\mu_0$ and variance $\sigma_0^2$,
      \textit{s}(\textit{t}) $\in N(\mu,\sigma^2)$ is a Gaussian signal with mean $\mu$ and variance $\sigma^2$. Let $\mu_1=\mu_0+\mu$ and 
      $\sigma_1^2=\sigma_0^2+\sigma^2$ since we assume that signal and noise are independent.   
      $H_{0}$ and $H_{1}$ are hypothesis 
      under which the signal of interest is either absent (hypothesis $H_{0}$) or present ($H_{1}$) along with the additive background noise n(t).
      As we demonstrate further, the proposed utilization of algorithm known as Cumulative Sum (CUSUM) introduced by \citet{b37,b28} allows for 
      mean only parameter to be sufficient when building a detector.

Changes detection scheme, used ino transient identification algorithm, is based on two generic detection procedures: log likelihood ratio (LLR) and CUSUM.

LLR test is based on using a ratio of two $pdf$ to build an indicator upon which a threshold can be applied. 
If the ratio exceeds given threshold, it indicates the prevalence of one of the hypothesis over another.

LR for a Gauss-distributed data using the model given in (\ref{eq2})  is:
\begin{equation}\label{eq4}
\ell (x)=\frac{pdf_{H_{1} } (x)}{pdf_{H_{0} } (x)} =\frac{\sigma _{0} }{\sigma _{1} } e^{\frac{-(x-\mu _{1} )^{2} }{2\sigma _{1}^{2} } } e^{\frac{(x-\mu _{0} )^{2} }{2\sigma _{0}^{2} } }
\end{equation}
We consider identification successful when $\ell(x)>\eta$, where $\eta$ is a predefined threshold for identification.

LLR is obtained by taking logarithm of (\ref{eq4}):
\begin{equation}\label{eq5}
\ln (\ell (x))=L(x)=\frac{1}{2} \ln \frac{\sigma _{0} }{\sigma _{1} } +\frac{(x-\mu _{0} )^{2} }{2\sigma _{0}^{2} } -\frac{(x-\mu _{1} )^{2} }{2\sigma _{1}^{2} } 
\end{equation}
Equation (\ref{eq5}) can now be expressed through either mean- or variance-based detector.

In the latter case, assuming equal and zero mean, we obtain:
\begin{equation}\label{eq6}
L(x)=\frac{1}{2} \ln \frac{\sigma _{0} }{\sigma _{1} } +\frac{x^{2} (\sigma _{1}^{2} -\sigma _{0}^{2} )}{2\sigma _{0}^{2} \sigma _{1}^{2} } 
\end{equation}

Assuming equal and unity variance, (\ref{eq5})  becomes: $L(x)=(\mu _{1} -\mu _{0} )(x-(\mu _{0} +\mu _{1} )/2)$.

Thus, defining
\begin{equation}\label{eq13}
\mu' =\frac{\mu _{0} +\mu _{1} }{2}
\end{equation}
and rescaling the log likelihood expression by the proportionality value ($\mu_1-\mu_0$) to obtain a threshold value.

 If the probability of Type I error $\alpha$ is fixed, the detection threshold can be expressed as $\eta =\frac{\sigma _{1} }{\sqrt{n} \Phi (1-\alpha )} +\mu _{1}$.

CUSUM  is a detection procedure proposed by \citet{b37,b28}. CUSUM is a repeated LLR test for a change from one known distribution to 
another. We assume that for known densities $f_0$ and $f_1$ there exists an unknown change point $v$, where the input sequence $X(n) \in f_0$ if $n<v$ and 
$X(n) \in f_1$ if $n>=v$. The CUSUM statistics is the one that satisfies the following recursion:
\begin{equation}\label{eq19}
W_{n} =\max \{ 0,W_{n-1} +L_{n} \}
\end{equation}
Where $L$ is a likelihood ratio for $f_0/f_1$ evaluated at $X(n)$. The procedure raises an alarm at time $t=\underset{n}{\operatorname{argmin}} \{ |W_{n}| > |h|\}$ for 
threshold $h$, which is discussed later.

If likelihood value from (\ref{eq5}) used with regard to variances, the CUSUM recursion would take the following form, based on the usual 
radiometric output being the averaged sum of the samples' squares or the energy detector:
\begin{equation}\label{eq21}
W_{n} =\max \{ 0,W_{n-1} +X_{_{n} }^{2} -r\}
\end{equation}
with the parameter $r$ denoting variance-based coefficient $r=\frac{2\cdot \ln (\frac{\sigma _{1} }{\sigma _{2} } )\sigma _{1}^{2} \sigma _{2}^{2} }{\sigma _{2}^{2} -\sigma _{1}^{2} }$. Since it relies on the full knowledge of variances of the signal and noise, it is not always practical to use CUSUM algorithm with the 
recursion denoted in (\ref{eq21}). 

Applying (\ref{eq13}) CUSUM may be expressed as a procedure, which starts with $W_0=0$, it recursively calculates:
\begin{equation}\label{eq23}
W_{n} =\max \{ 0,W_{n-1} +X_{n} -\mu' \}
\end{equation}
and stops as soon as $W_n$ exceeds threshold $h$.

Generally speaking, equation (\ref{eq23}) can be rewritten as $W_{n} =\max \{ 0,W_{n-1} +X_{n} -r\}$, where $r$ being a configurable parameter. 
Choice of this parameter is ruled by the expected behavior of the procedure, as stated by \citet{b37}: 
``Scoring is chosen so that the mean sample path on the chart when quality is satisfactory is downwards \dots  and is upwards when quality 
is unsatisfactory''. As demonstrated by \citet{b9} the optimal value for parameter r will be the largest acceptable mean value or the 
smallest unacceptable mean value.

One of the benefits of CUSUM scheme for signal detection is its stability in the presence of regression behavior in input data \citep{b19,b9}, which is
 due to the gradual changes in the system temperature or changes of the overall sky temperature for large field of view or ``all-sky'' telescopes.

\section{Algorithm}

Choice between mean-based and variance-based CUSUM schemes depends on the type of the receiver used. For a linear output receiver, sampled according to Nyquist, the output represents voltage changes with time. Both mean- and variance-based indicator function can
be used. However, mean-based indicator is preferred because it does not rely on variance-based separation of noise and signal. For full power square-law detector, however,  mean-based indicator function becomes unusable due to the loss of information on mean. Therefore, variance-based CUSUM
scheme should be used discounting the utilization of  mean \citep{b49}.

CUSUM scheme requires a threshold value $h$ to be chosen, which, when crossed, identifies the point of abrupt change in the statistical characteristics of the signal.
Threshold value $h$ can be derived from Wald test on the mean of a normal population \citep{b50}. Assuming that $\mu_1>\mu_0$, the value of threshold $h$ is equivalent to a sequence of Wald 
sequential tests \citep{b51} with boundaries $(0,h)$:
\begin{equation}\label{eq_threshold}
h=-\frac{\sigma_1^2}{\sigma_0^2}\frac{\ln{\alpha}}{(\mu_1-\mu_0)}
\end{equation}
where $\alpha$ was interpreted by \citet{b52} as a crude approximation to the proportion of samples that trigger false alarm. This value can also
be interpreted as a probability of false alarm in a traditional sense. Since we assumed that $\sigma_1=\sigma_0$ (\ref{eq_threshold}) can be 
rewritten as $h=-\frac{\ln{\alpha}}{(\mu_1-\mu_0)}$. 

\begin{figure}
\centering
\includegraphics[width=165mm, height=36.1mm]{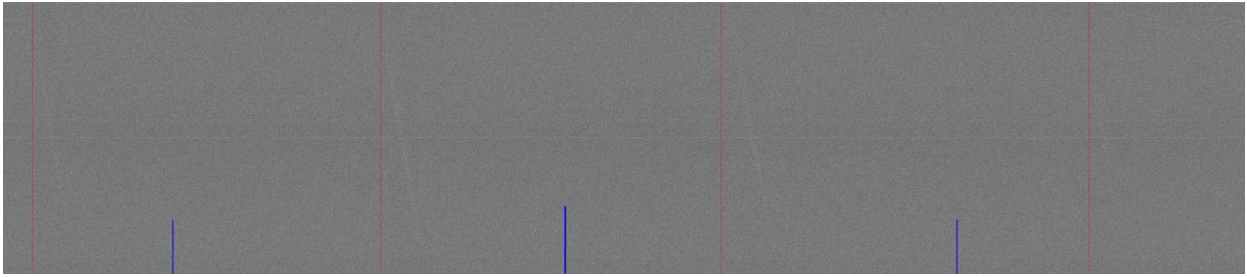}
\caption{Time/frequency representation of pulsar $J0835-4510$, observed at Parks telescope, central frequency $1416$ MHz, bandwidth $64$ MHz, recorded with 2-bit VLBI recorder.  
   The recording started at  $2009.12.10 - 17:25:51 UTC$. Short bars mark  $0.1 s$ intervals. Vertical red lines mark the beginning of detected pulses. }
\label{fig:vela_det}
\end{figure}

The algorithm was tested on the data obtained from Parks radiotelescope. Input data contained 1 second observation from
the Vela pulsar PSR J0835-4510. The observation was sampled at 1416 MHz with 64 MHz bandwidth. PSR J0835-4510 has a period of 89 ms. Figure~\ref{fig:vela_det}
represents a portion of the observed data with the detected pulses marked by vertical red lines, with blue lines representing 0.1 second intervals. Out of 10 
impulses contained in the test data, 9 were identified.

90\% of pulses being correctly identifed prove the applicability of CUSUM procedure described above upon the raw, time-domain
radio data.  Detectability of a signal is controled by and relied upon the preset threshold, which
is calculated based on the assumed probability distribution of data (\ref{eq_threshold}). 

\section{Conclusion}

We have presented the algorithm, which provides reliable and computationally efficient detection of dispersed radio transients in time domain.
The algorithm is well-suited for being implemented on FPGA or GPU platforms for real time detection on arrays such as SKA. Statistical methods used in the algorithm provide easy staging of detection process across multiple handling points giving the opportunity for significant reduction of data volume at each consequent stage of detection.

\section*{Acknowledgments}

The authors would like to thank Aidan Hotan of Curtin Institute of Radioastronomy for making $J0835-4510$ pulsar data available to us.

\end{document}